\documentclass[fleqn,twoside]{article}
\usepackage{espcrc2}

\usepackage{graphicx}
\usepackage[figuresright]{rotating}

\newcommand{\AmS}{{\protect\the\textfont2
  A\kern-.1667em\lower.5ex\hbox{M}\kern-.125emS}}

\newcommand{\lwig}{\mbox{\ \raisebox{.3ex}
    {$<$}$\!\!\!\!\!$\raisebox{-.9ex}{$\sim$}\ }}
\newcommand{\gwig}{\mbox{\ \raisebox{.3ex}
    {$>$}$\!\!\!\!\!$\raisebox{-.9ex}{$\sim$}}\ }
\newcommand{\lambdabar}{{\hbox{$\lambda_e$\kern-1.9ex\raise+0.45ex\hbox{--}
\kern+0.2ex}}}

\title{\vbox to 0pt{{\vss\flushright\normalsize\rm
DESY 04-180
\endflushright}}
Extremely energetic cosmic neutrinos and their 
impact on particle physics and cosmology}

\author{A. Ringwald\address[DESY]{Deutsches Elektronen-Synchrotron DESY, \\ 
        Notkestra\ss e 85, D--22607 Hamburg, Germany}%
        \thanks{Invited talk at Cosmic Ray International Seminar 2004, Catania, Italy.}
       }
       
\begin{document}

\begin{abstract}
Existing and planned neutrino detectors, sensitive in the
energy regime from $10^{17}$~eV to $10^{23}$~eV, offer opportunities for 
particle physics and cosmology. In this contribution, we discuss particularly  
the possibilities to infer information about physics beyond the Standard Model
at center-of-mass energies beyond the reach of the Large Hadron Collider, 
as well as to detect big bang relic neutrinos via absorption dips in the observed neutrino spectra. 
\vspace{1pc}
\end{abstract}

\maketitle

\section{INTRODUCTION}

\begin{figure}
\vspace{-6ex}
\begin{center}
\includegraphics*[bbllx=20pt,bblly=221pt,bburx=572pt,bbury=608pt,width=7.5cm]{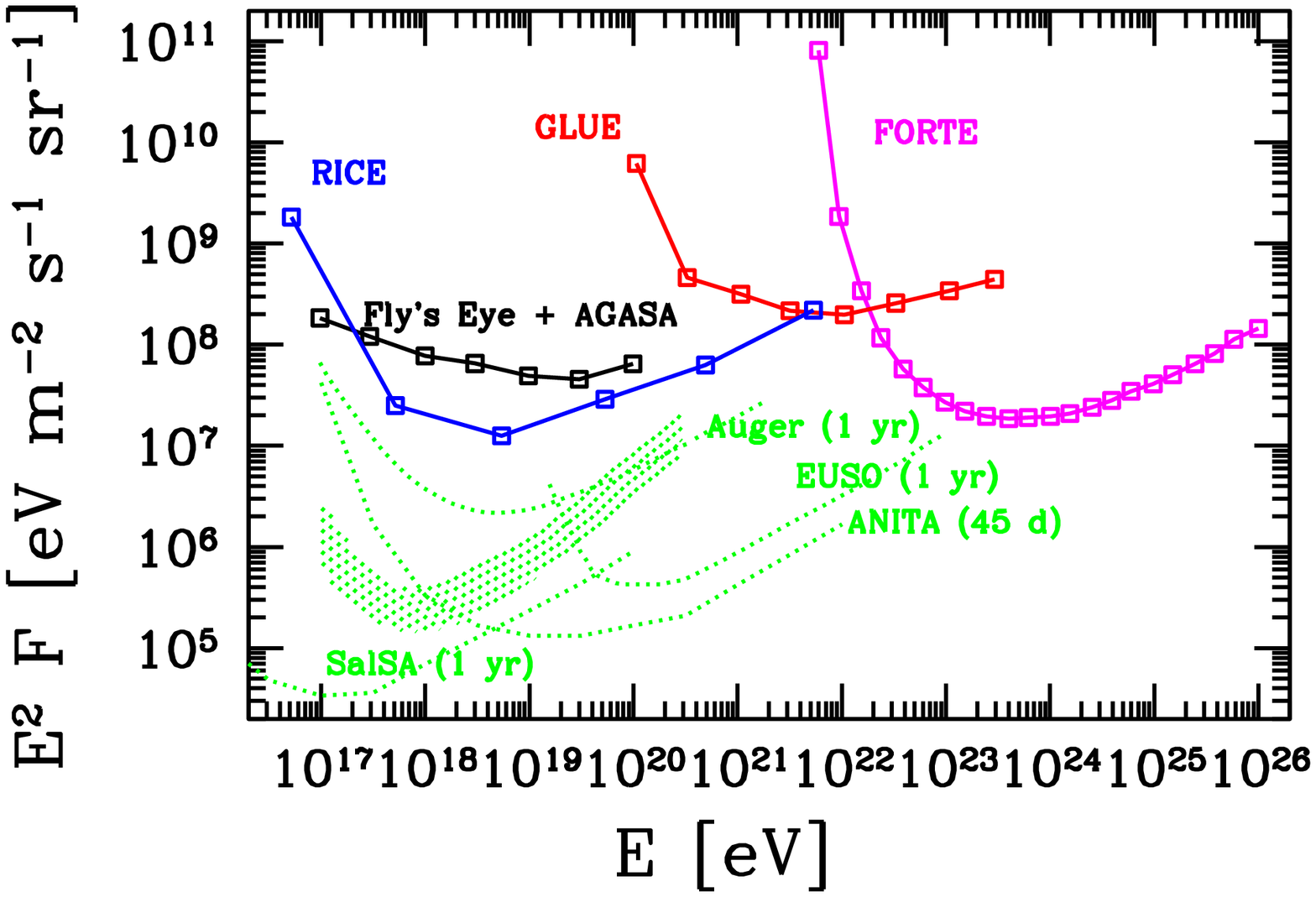}
\includegraphics*[bbllx=20pt,bblly=221pt,bburx=572pt,bbury=608pt,width=7.5cm]{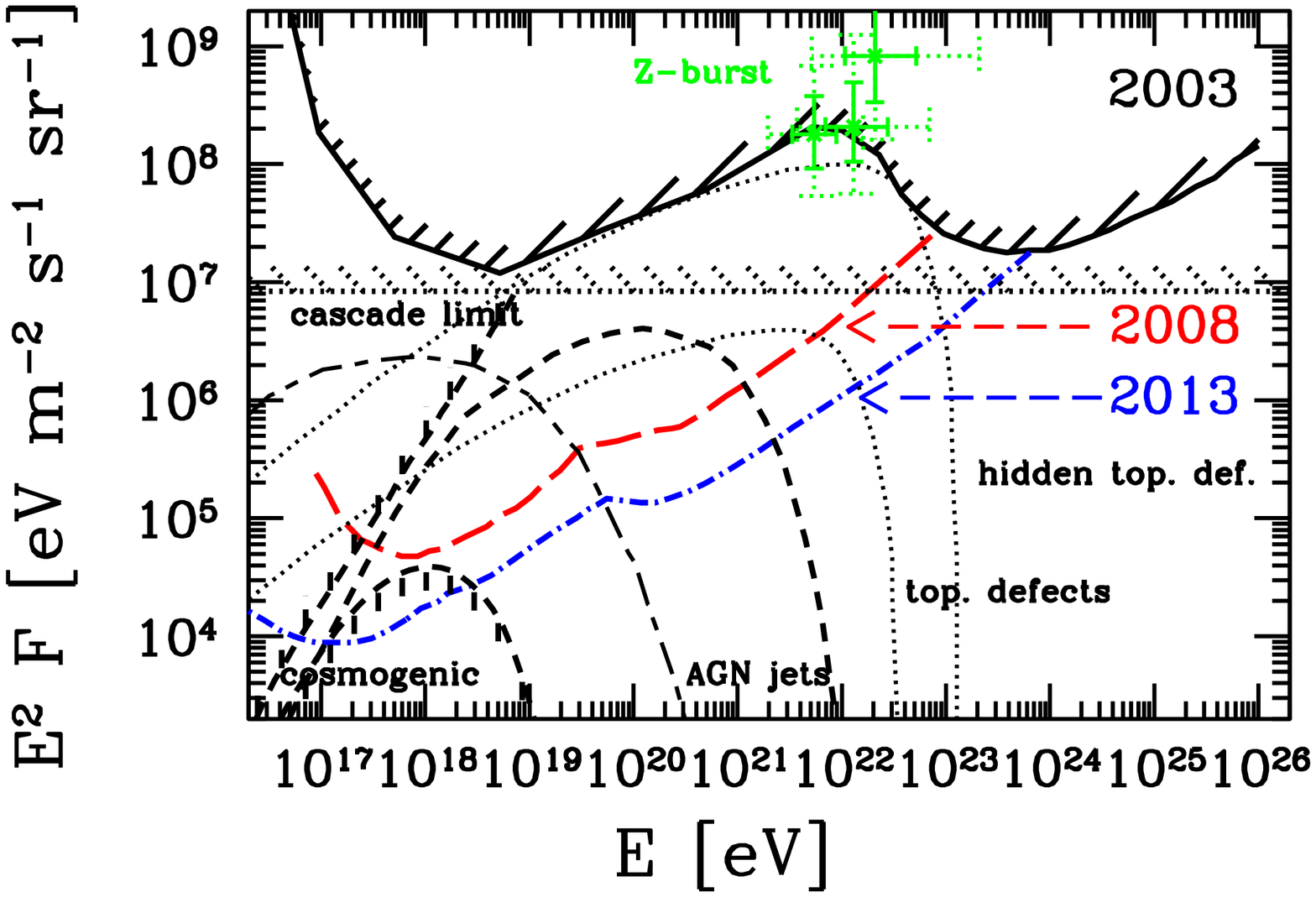}
\vspace{-8ex}
\caption[...]{Current status and next decade prospects for EHEC$\nu$ physics, 
expressed in terms of diffuse neutrino fluxes per flavor, 
$F_{\nu_\alpha}+F_{\bar\nu_\alpha}$, $\alpha =e,\mu,\tau$~\cite{Eberle:2004ua}.\\
{\em Top:} Upper limits from  
RICE~\cite{Kravchenko:2003tc}, GLUE~\cite{Gorham:2003da},  
FORTE~\cite{Lehtinen:2003xv}, and Fly's Eye~\cite{Baltrusaitis:mt} and 
AGASA~\cite{Yoshida:2001,Anchordoqui:2002vb}. 
Also shown are projected sensitivities 
of Auger in $\nu_e$, $\nu_\mu$ modes and in $\nu_\tau$ mode 
(bottom swath)~\cite{Bertou:2001vm}, ANITA~\cite{Gorham:Anita}, EUSO~\cite{Bottai:2003i}, 
and SalSA~\cite{Gorham:2001wr}, corresponding to 
one event per energy decade and indicated duration.\\  
{\em Bottom:} Roadmap for improvement 
in the next decade (2008 and 2013), 
corresponding to one event per energy decade, as well as the current (2003)
observational upper bound (solid-shaded) obtained from Fig.~\ref{roadmap} (top). 
Also shown is a wide sample of predictions for EHEC$\nu$ fluxes.
\label{roadmap}} 
\end{center}
\end{figure}

Existing observatories for extremely high energy cosmic neutrinos 
(EHEC$\nu$), such as RICE~\cite{Kravchenko:2003tc}, GLUE~\cite{Gorham:2003da}, 
and FORTE~\cite{Lehtinen:2003xv}, 
have recently put sensible upper limits on the neutrino flux in the energy region
from $10^{17}$~eV to $10^{26}$~eV (cf. Fig.~\ref{roadmap}). 
Furthermore, recent proposals for progressively larger EHEC$\nu$ detectors, such as 
the Pierre Auger Observatory~\cite{Bertou:2001vm}, IceCube~\cite{Ahrens:2002dv}, 
ANITA~\cite{Gorham:Anita}, EUSO~\cite{Bottai:2003i}, SalSA~\cite{Gorham:2001wr}, 
and OWL~\cite{Stecker:2004wt},  
together with conservative neutrino flux predictions, offer credible hope 
that the collection of an appreciable event sample above $10^{17}$~eV may be realized within
this decade~\cite{Spiering:2003xm} (cf. Fig.~\ref{roadmap}). 
This will provide an opportunity for particle physics beyond the reach of 
the Large Hadron Collider (LHC). 
There is even a remote possibility of a sizeable event 
sample above $10^{21}$~eV. If the corresponding more speculative neutrino fluxes are 
realized in nature, EHEC$\nu$ open a window to cosmology:
it may be possible to detect the  cosmic neutrino background 
(C$\nu$B) via absorption features in neutrino spectra. 

In this contribution, we will have a closer look at these exciting possibilities. 

\section{\boldmath EHEC$\nu$ AND PHYSICS BEYOND THE STANDARD MODEL}

\begin{figure}
\vspace{-2.7ex}
\begin{center}
\includegraphics*[width=7.5cm]{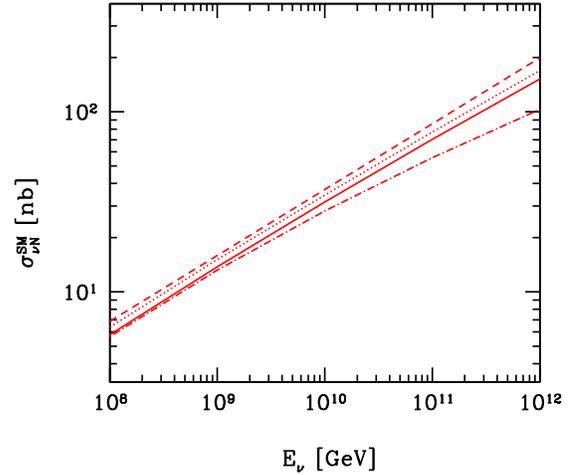}
\vspace{-12ex}
\caption[...]{Standard Model $\nu N$ total cross section $\sigma_{\nu N}^{\rm tot}$ 
at extremely high neutrino energies $E_\nu$ obtained by various  
perturbative QCD resummation techniques~\cite{Tu:2004ms}:
from~\cite{Kwiecinski:1998yf} (solid), 
from~\cite{Gandhi:1998ri} (dotted), 
from~\cite{Gluck:1998js} (dashed), and from 
~\cite{Kutak:2003bd} (dashed-dotted).  
\label{SM}} 
\end{center}
\end{figure}

Cosmic neutrinos with energies $E_\nu$ above $10^{17}$~eV probe  
neutrino-nucleon scattering at center-of-mass (c.m.) energies above 
\begin{equation} 
\sqrt{s_{\nu N}}\equiv \sqrt{2m_NE_\nu} \simeq 14\ 
\left( \frac{E_\nu}{10^{17}\ {\rm eV}}\right)^{1/2}
\ {\rm TeV}, 
\end{equation}
beyond the proton-proton c.m. energy  $\sqrt{s_{pp}}=14$~TeV of the LHC, 
and Bjorken-$x$ values below
\begin{eqnarray}
x &\equiv & \frac{Q^2}{y\, s_{\nu N}} 
\\ \nonumber 
&\simeq & 2\times 10^{-4}\  
\left( \frac{Q^2}{m_W^2}\right)  
\left( \frac{0.2}{y}\right) 
\left( \frac{10^{17}\ {\rm eV}}{E_\nu}\right)
\,,
\end{eqnarray}
where $Q^2$ is the momentum transfer squared, $m_W\simeq 80$~GeV the $W$-boson mass, 
and $y$ the inelasticity parameter. 
Under these kinematical conditions, the predictions for $\nu N$ scattering 
from the perturbative Standard Model (SM) are quite safely under control (cf. Fig.~\ref{SM}), 
notably thanks to the input from measurements of deep-inelastic $ep$ scattering at 
HERA (e.g.,~\cite{Adloff:2003uh,Chekanov:2003vw}). 
This makes it possible to search for enhancements in the $\nu N$ cross section 
due to physics beyond the (perturbative) SM, such as, e.g.,   
electroweak sphaleron production (non-perturbative $B+L$ violating 
processes)~\cite{Aoyama:1986ej,Ringwald:1989ee,Espinosa:qn,Khoze:1991mx}, and 
Kaluza-Klein, black hole, $p$-brane or string ball production in TeV scale gravity 
models~\cite{Antoniadis:1990ew,Lykken:1996fj,Arkani-Hamed:1998rs,Randall:1999ee}. 

\begin{figure}
\begin{center}
\includegraphics*[bbllx=20pt,bblly=221pt,bburx=572pt,bbury=608pt,width=7.5cm]{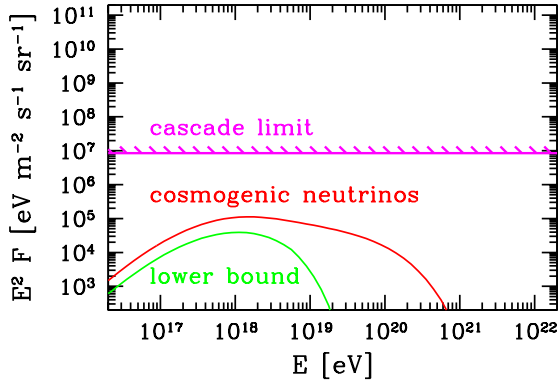}
\vspace{-10ex}
\caption[...]{Predicted EHEC$\nu$ fluxes per flavour, $F_{\nu_\alpha}+F_{\bar\nu_\alpha}$, 
$\alpha =e,\mu,\tau$~\cite{Tu:2004ms}: (i) Best fit and lower bound cosmogenic neutrino flux (solid) 
under the assumption that the observed EHECR below $10^{20}$~eV are protons from uniformly 
distributed extragalactic sources~\cite{Fodor:2003ph}; 
(ii) Cascade limit~\cite{Berezinsky:1975} (shaded) 
on transparent neutrino sources from~\cite{Mannheim:1998wp}.
\label{fluxes}} 
\end{center}
\end{figure}

Of central importance in the evaluation of the prospects of EHEC$\nu$ for physics
beyond the SM is their expected flux $F_\nu$ to which we turn our 
attention next. 
Though atmospheric neutrinos, i.\,e. neutrinos produced in hadronic showers in the atmosphere, 
are certainly present, their flux in the energy region of interest is 
negligible. 
Much more promising, but also more or less guaranteed are the so-called 
cosmogenic neutrinos which are produced when extremely high energy cosmic rays (EHECR), 
notably protons or even heavy nuclei, inelastically scatter off the cosmic microwave background (CMB) 
radiation~\cite{Greisen:1966jv,Zatsepin:1966jv}
in processes of the type 
$p\,\gamma\to N\,\pi's\to N\,\nu's$~\cite{Beresinsky:1969qj,Beresinsky:1970,Stecker:1979ah}. 
Recent estimates of these fluxes  can be found in 
Refs.~\cite{Yoshida:1993pt,Protheroe:1996ft,Engel:2001hd,Kalashev:2002kx,%
Fodor:2003ph,Semikoz:2003wv,Hooper:2004jc,Anchordoqui:2004}
(cf. Figs.~\ref{roadmap} (bottom) and~\ref{fluxes}).  

Whereas the cosmogenic neutrino flux represents a reasonable lower limit on the 
ultrahigh energy neutrino flux, it is also useful to have an upper limit on the 
latter. The least model dependent is the cascade limit~\cite{Berezinsky:1975} 
on transparent  neutrino sources (cf. Figs.~\ref{roadmap} (bottom) and~\ref{fluxes}). 
It applies to all scenarios where neutrinos originate from pion decays. 
These neutrinos are accompanied by photons 
and electrons which cascade down in energy during their propagation through the universe. 
The cascade limit arises from the requirement that 
the associated diffuse gamma-ray fluxes should not exceed measurements\footnote{The cascade 
limit shown in Fig.~\ref{fluxes} from~\cite{Mannheim:1998wp} exploits the 
measurement of the diffuse $\gamma$ ray background from 30 MeV to 100 GeV by EGRET~\cite{Sreekumar:1997un}.
A lower extragalactic contribution to the $\gamma$ ray 
background than that inferred in~\cite{Sreekumar:1997un}, by roughly a factor of two, 
has been proposed recently~\cite{Strong:2003ex,Keshet:2003xc}.}. 

\begin{figure}
\begin{center}
\includegraphics*[bbllx=25pt,bblly=230pt,bburx=584pt,bbury=610pt,width=7.5cm]{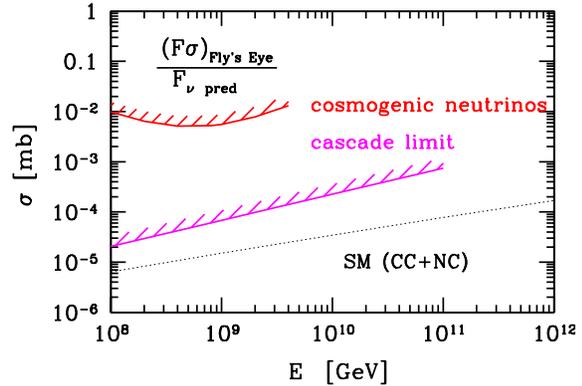}
\vspace{-8ex}
\caption[...]{Upper limits on the $\nu N$ cross section (shaded) from the non-observation of
deeply-penetrating showers by Fly's Eye~\cite{Baltrusaitis:mt}, 
for various assumptions about the 
EHEC$\nu$ flux (cf. Fig.~\ref{fluxes})~\cite{Tu:2004ms} (updated from~\cite{Morris:1993wg}). 
Also shown is the SM total 
(charged current (CC) plus neutral current (NC)) cross section from~\cite{Gandhi:1998ri} (dotted). 
It is implicitely assumed that $\sigma_{\nu N}\lwig 0.01$--$0.5$~mb and that all energy is visible.  
\label{mod-indep-lim-fe}} 
\end{center}
\end{figure}

\begin{figure}
\begin{center}
\includegraphics*[bbllx=25pt,bblly=172pt,bburx=587pt,bbury=588pt,width=7.5cm]{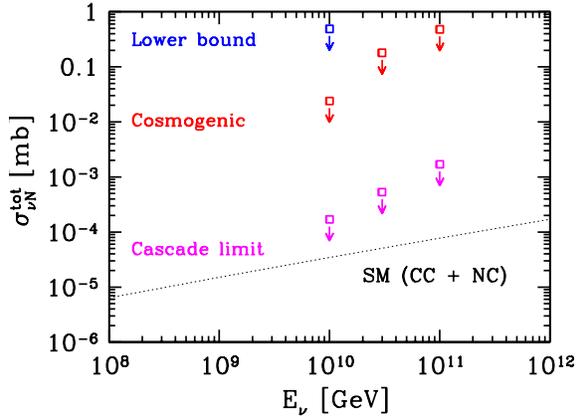}
\vspace{-8ex}
\caption[...]{Upper limits on the $\nu N$ cross section (shaded) from the non-observation of
deeply-penetrating showers by AGASA~\cite{Yoshida:2001}, for various assumptions about the 
EHEC$\nu$ flux (cf. Fig.~\ref{fluxes})~\cite{Tu:2004ms}. 
It is implicitely assumed that $\sigma_{\nu N}\lwig 0.01$--$0.5$~mb and that all energy is visible.  
\label{mod-indep-lim-agasa}} 
\end{center}
\end{figure}

Since the rate of neutrino-initiated showers is proportional to integrated flux
times cross section, 
\begin{equation}
      R  
      \propto  
      \int {\rm d}E_\nu\,{F_\nu (E_\nu )}\,\sigma_{\nu N}(E_\nu )
\,,
\end{equation}
the non-observation of quasi-horizontal or deeply-penetrating neutrino-induced air showers
as reported by, e.g., Fly's Eye~\cite{Baltrusaitis:mt},  AGASA~\cite{Yoshida:2001}, 
and RICE~\cite{Kravchenko:2003tc} 
can be turned into an upper bound on the neutrino-nucleon cross section if a certain 
prediction for the neutrino flux is exploited~\cite{Berezinsky:kz}, see, e.g., 
Refs.~\cite{Morris:1993wg,Tyler:2000gt,Ringwald:2001vk,Anchordoqui:2001cg,%
Anchordoqui:2003jr,Anchordoqui:2004}.  
This is exemplified in Figs.~\ref{mod-indep-lim-fe} and \ref{mod-indep-lim-agasa}, which display
the limits on $\sigma_{\nu N}$ from the Fly's Eye~\cite{Baltrusaitis:mt}  
and AGASA~\cite{Yoshida:2001} constraints on deeply-penetrating showers, respectively,  
for various assumptions about the EHEC$\nu$ flux. These bounds are considerably higher
than the SM cross section, albeit in the post-LHC energy region.   
Clearly, the most conservative upper bound arises from the most conservative
assumption on the EHEC$\nu$ flux. If the latter is as low as the lower bound on the cosmogenic
neutrino flux in Fig.~\ref{fluxes}, the upper bound on $\sigma_{\nu N}$ is nearly 
non-existent, notably because in such an analysis it is implicitely assumed that 
neutrinos are indeed deeply-penetrating, corresponding to  
$\sigma_{\nu N}\lwig 0.01$--$0.5$~mb. 

These model-independent bounds can be improved if a particular particle physics scenario
is probed. We shall consider here  electroweak sphaleron production and black hole production.  

\begin{figure}
\begin{center}
\includegraphics*[bbllx=32pt,bblly=124pt,bburx=584pt,bbury=610pt,width=7.5cm]{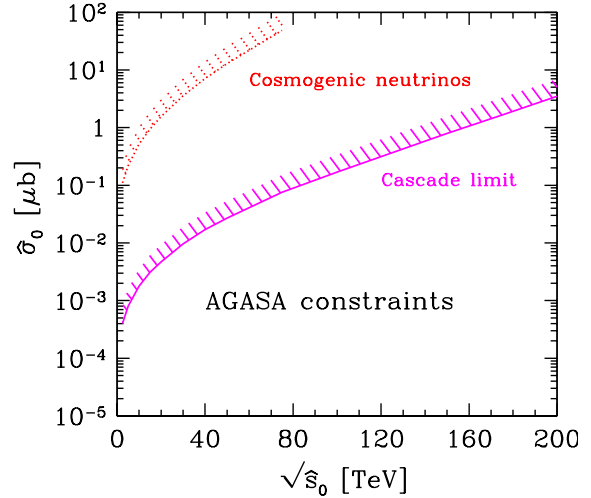}
\vspace{-8ex}
\caption[...]{Upper limit on the parton-level sphaleron production cross section 
$\hat\sigma_0$ as a function of threshold energy $\sqrt{\hat s_0}$ 
(cf. Eq.~(\ref{sphaleron-cross})) from the non-observation of deeply-penetrating 
showers by AGASA~\cite{Yoshida:2001}, exploiting various EHEC$\nu$ predictions 
(cf.~Fig.~\ref{fluxes})~\cite{Tu:2004ms} (updated from~\cite{Morris:1993wg}).
\label{sphaleron-lim}} 
\end{center}
\end{figure}

It is well known~\cite{Aoyama:1986ej,Ringwald:1989ee,Espinosa:qn,Khoze:1991mx} 
that the cross section for non-perturbative electroweak sphaleron production---and associated
$B+L$ violation, as well as multi-$W$ and -$Z$ production---is 
exponentially small, $\hat\sigma^{\rm sp}\ll 10^{-100}$~pb, albeit 
exponentially growing, at parton-parton c.m. energies $\sqrt{\hat s}\ll m_{\rm sp} \equiv  
\pi m_W/\alpha_W\simeq 7.5$~TeV,
where $\alpha_W\simeq 0.03$ is the electroweak fine structure constant. 
The fate of these processes at $\sqrt{\hat s}\gwig \pi m_W/\alpha_W$, 
notably the level at which the exponential growth of the cross section finally saturates, 
is, however, uncertain and there exist only estimates, educated guesses and bounds in this
energy domain (e.g.,~\cite{Ringwald:2002sw,Bezrukov:2003er,Ringwald:2003ns}). 
Therefore, it is of considerable interest to get information about these processes from 
EHEC$\nu$ physics (see also~\cite{Han:2003ru}). 
For such practical purposes, the electroweak sphaleron production
cross section, at parton level, may be parametrized by a step function~\cite{Morris:1993wg}, 
\begin{equation}
\label{sphaleron-cross}
\hat\sigma^{\rm sp} =\hat\sigma_0\ \theta (\sqrt{\hat s}-\sqrt{\hat s_0})
\,.
\end{equation}
As shown in Fig.~\ref{sphaleron-lim}, the AGASA constraints 
on deeply-penetrating showers give already sensible exclusion regions for the 
parton-level cross section $\hat\sigma^0$ and threshold energy $\sqrt{\hat s_0}$ 
at post-LHC energies, for reasonable assumptions about the EHEC$\nu$ flux. 
Again, the upper limits on $\hat\sigma^0$ disappear above $\gwig 100$~$\mu$b. 
We note here in passing, that, for even higher and more speculative cross sections,  
$\gwig 1$--$10$~mb, 
electroweak sphaleron production qualifies~\cite{Fodor:2003bn} as a particular 
strongly interacting neutrino scenario~\cite{Beresinsky:1969qj,Fodor:2004tr}, 
according to which the mysterious EHECR beyond the 
predicted Greisen-Zatsepin-Kuzmin (GZK) cutoff at 
$E_{\rm GZK}\simeq 4\times 10^{19}$~eV~\cite{Greisen:1966jv,Zatsepin:1966jv} 
are initiated by cosmogenic neutrinos.  

\begin{figure}
\begin{center}
\includegraphics*[bbllx=26pt,bblly=227pt,bburx=564pt,bbury=601pt,width=7.5cm]{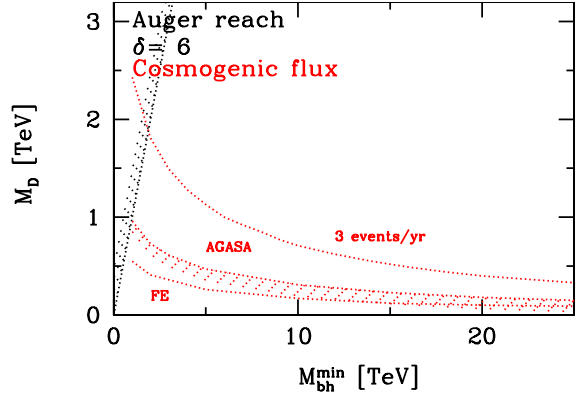}
\vspace{-8ex}
\caption[...]{Projected Auger reach in the black hole production parameters (fundamental 
Planck scale $M_D$ and minimum mass of produced black hole $M_{\rm bh}^{\rm min}$) 
for $\delta = 6$ extra 
dimensions~\cite{Tu:2004ms} (updated from~\cite{Ringwald:2001vk}),  
exploiting the cosmogenic neutrino flux from Fig.~\ref{fluxes}.
The shaded dotted lines indicate the constraints arising from 
the non-observation of horizontal showers by Fly's Eye and AGASA.  
\label{bh-auger}} 
\end{center}
\end{figure}

Such strongly interacting neutrino scenarios may also 
arise in TeV scale gravity models involving extra 
dimensions~\cite{Antoniadis:1990ew,Lykken:1996fj,Arkani-Hamed:1998rs,Randall:1999ee}. 
For example, neutrino-initiated $p$-brane production in models 
with warped extra dimensions~\cite{Ahn:2002mj,Jain:2002kf} or the production of 
resonances in low scale string unification models~\cite{Burgett:2004ac}
may lead to a viable solution of the GZK problem. But very often, the cross sections
turn out to be either not large enough or severely constrained at lower energies by the 
observational constraints on deeply-penetrating showers~\cite{Anchordoqui:2002it}. 
Let us concentrate here on microscopic black holes 
whose copious production in high energy collisions at c.m. energies above 
the fundamental Planck scale $M_D\gwig$~TeV is one of 
the least model-dependent predictions of TeV scale gravity 
scenarios~\cite{Argyres:1998qn,Banks:1999gd,Aref'eva:1999bm,Emparan:2000rs,Giddings:2000ay,Emparan:2001ce}. 
Correspondingly, the LHC may turn into a factory of black holes
at which their production and evaporation may be studied in 
detail~\cite{Giddings:2001bu,Dimopoulos:2001hw}. 
But even before the commissioning of the LHC, the first signs of black
hole production may be observed at EHEC$\nu$  
observatories~\cite{Feng:2002ib,Emparan:2001kf,Anchordoqui:2002ei,Ringwald:2001vk,Anchordoqui:2001cg,%
Kowalski:2002gb,Alvarez-Muniz:2002ga,Dutta:2002ca,Anchordoqui:2003jr}. Moreover, the constraints 
on black hole production from the non-observation of horizontal showers by Fly's Eye 
and AGASA turn out~\cite{Ringwald:2001vk,Anchordoqui:2001cg,Anchordoqui:2003jr} 
to be 
competitive with other currently available constraints on TeV-scale gravity 
which are mainly based on interactions associated with
Kaluza-Klein gravitons, 
according to which a fundamental Planck scale as low as 
$M_D = {\mathcal O}(1)$ TeV is still allowed for $\delta\geq 6$ flat 
or $\delta\geq 1$ warped extra dimensions~\cite{Eidelman:2004wy} 
(cf. Fig.~\ref{bh-auger}). 

We have emphasized here the current constraints from EHEC$\nu$ on physics beyond the SM. 
A more detailed account of the particle physics reach of the planned EHEC$\nu$ observatories 
can be found in~\cite{Tu:2004ms,Han:2004kq}. 

\begin{figure}
\begin{center}
\includegraphics*[bbllx=20pt,bblly=227pt,bburx=587pt,bbury=606pt,width=7.5cm]{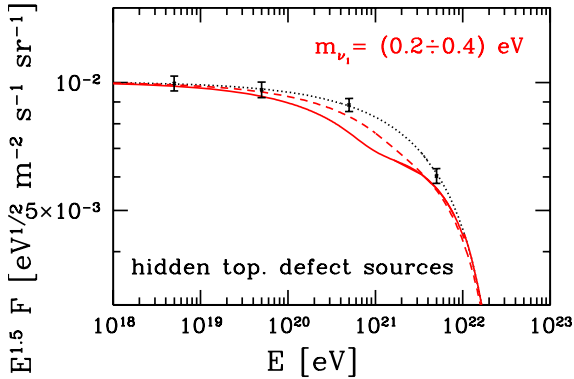}
\vspace{-8ex}
\caption[...]{Predicted neutrino flux at Earth, summed over all flavors, 
from hidden-sector topological defects 
(cf.~Fig.~\ref{roadmap} (bottom))~\cite{Eberle:2004ua}. 
Curves are without (dotted) and with relic neutrino absorption.  
Assumed neutrino masses are degenerate at $m_{\nu}=0.2$~eV (dashed) and $m_{\nu}=0.4$~eV (solid).
The error bars indicate the statistical accuracy achievable 
per energy decade by the year 2013, for a flux which saturates today's observational bound from 
Fig.~\ref{roadmap} (bottom)---which is also sufficient 
to explain the EHECR above $E_{\rm GZK}$ via the $Z$-burst mechanism
(cf. Fig.~\ref{Zburst-moderate}). 
\label{dip-hidden}} 
\end{center}
\end{figure}

\section{\boldmath RELIC NEUTRINO AB\-SORPTION SPECTROS\-CO\-PY}

Neutrinos are the elementary particles with the weakest known interactions. 
Correspondingly, they can propagate to us through the CMB and C$\nu$B 
without significant energy loss even from cosmological distances. 
A possible exception to this transparency 
is resonant annihilation of EHEC$\nu$ on big-bang relic anti-neutrinos (and vice versa) into 
$Z$-bosons~\cite{Weiler:1982qy,Weiler:1983xx,Gondolo:1991rn,Roulet:1992pz,Yoshida:1996ie}. 
This occurs near the respective resonance 
energies,
\begin{eqnarray}
\label{Eres}
E_{\nu_i}^{\rm res} = \frac{m_Z^2}{2\,m_{\nu_i}} \simeq 4\times 10^{21}  
\left( \frac{{\rm eV}}{m_{\nu_i}}\right)\ {\rm eV}
\,, 
\end{eqnarray}
with $m_Z\simeq 91$~GeV denoting the mass of the $Z$-boson 
and $m_{\nu_i}$ ($i=1,2,3$) the non-zero neutrino masses 
-- for which there is rather convincing evidence inferred from the apparent observation of neutrino 
oscillations~\cite{Eidelman:2004wy}. On resonance, the corresponding cross sections are enhanced
by several orders of magnitude.  This leads to 
a few percent probability of annihilation within the Hubble radius of the universe, 
even if one neglects further enhancing effects due to cosmic evolution. 
Indeed, it appears that 
-- apart from the indirect evidence 
to be gained from cosmology, e.g., big-bang nucleosynthesis and large-scale structure 
formation -- this annihilation mechanism 
is the unique process having sensitivity to the C$\nu$B~\cite{Weiler:1982qy}.
Moreover, observation of the absorption dips would present one of the few opportunities to 
determine absolute neutrino masses~\cite{Paes:2001nd,Bilenky:2002aw}.    

Apart from the absorption features in the EHEC$\nu$ spectra, 
other signatures of annihilation are emission 
features~\cite{Fargion:1999ft,Weiler:1999sh,Yoshida:1998it,Fodor:2001qy,Fodor:2002hy} ($Z$-bursts) 
as protons (or photons) with energies spanning a decade or more above the GZK cutoff. 
The association of $Z$-bursts with 
the mysterious EHECR observed above $E_{\rm GZK}$ is a controversial 
possibility~\cite{Fargion:1999ft,Weiler:1999sh,Yoshida:1998it,Fodor:2001qy,Fodor:2002hy,%
Kalashev:2001sh,Gorbunov:2002nb,Semikoz:2003wv,Gelmini:2004zb}. 

\begin{figure}
\begin{center}
\includegraphics*[bbllx=27pt,bblly=229pt,bburx=564pt,bbury=604pt,width=7.5cm]{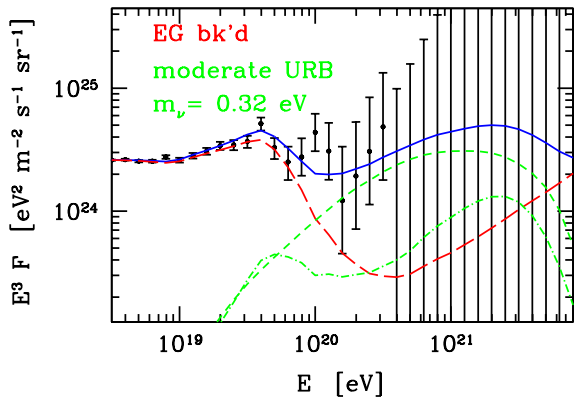}
\vspace{-8ex}
\caption[...]{Combined EHECR data with their error bars
and the best fit from $Z$-bursts (solid line), corresponding to the sum of 
background protons (long-dashed), $Z$-burst protons (dash-dotted)
and $Z$-burst photons (short-dashed)~\cite{Fodor:2002hy}. The necessary neutrino
flux is close to the current observational upper bound (cf.~Fig.~\ref{roadmap}). 
\label{Zburst-moderate}} 
\end{center}
\end{figure}

The possibility to confirm the existence of relic neutrinos within 
the next decade from a measurement 
of the aforementioned absorption dips in the EHEC$\nu$ 
flux was recently investigated in~\cite{Eberle:2004ua}. 
The presently planned neutrino observatories (cf. Fig.~\ref{roadmap}),
operating in the energy regime above $10^{21} \ {\rm eV}$, appear to 
be sensitive enough 
to lead us, within the next decade, into an era of relic neutrino absorption 
spectroscopy, provided that the neutrino mass
is sufficiently large, $m_\nu\gwig\, 0.1 \ {\rm eV}$ and the flux of the EHEC$\nu$ 
at the resonant energies is close to current observational 
bounds (cf. Fig.~\ref{dip-hidden}). 
In this case, the associated $Z$-bursts 
must also be seen as post-GZK events at the planned EHECR detectors (cf. Fig.~\ref{Zburst-moderate}). 
Relic neutrino overdensities in galaxy clusters within the local GZK zone, such as Virgo,   
may allow to search for directional dependences in the post-GZK emission 
events~\cite{Singh:2002de,Ringwald:2004np}.   

\section{CONCLUSIONS}

We have reviewed some particle physics and cosmology opportunities of EHEC$\nu$ observatories.

We have shown that already now EHEC$\nu$ data imply sensible constraints on post-LHC enhancements
in the neutrino-nucleon cross section. Clearly, a discovery 
of new physics exploiting EHEC$\nu$ within the next decade needs large deviations from the SM. 

As far as relic neutrino absorption spectroscopy is concerned, we have emphasized
that a detection of the absorption dips within the next decade needs an extraordinary large
EHEC$\nu$ flux close to the current observational limit and a quasi-degenerate neutrino
mass spectrum, $m_\nu\gwig 0.1$~eV. 

EHEC$\nu$ physics will be exciting in the next decade!

\section*{ACKNOWLEDGMENTS}

I would like to thank Luis Anchordoqui, Birgit Eberle, Zoltan Fodor, Sandor Katz, Marek Kowalski, 
Liguo Song, Huitzu Tu, Tom Weiler, and Yvonne Wong for the nice collaboration.

\end{document}